\documentclass{iopart}

	\usepackage{iopams}
	\usepackage{setstack}
	\usepackage{graphicx}
	\usepackage{subfigure}
	\usepackage{mcite}
	\usepackage{dsfont}
	\usepackage{mathrsfs}
	\usepackage{hyperref}
\newcommand{\vv}{\upsilon}

\newcommand{\X}{\tilde{X}}

\begin{document}

\title[Phase Space Geometry and Chaotic Attractors]{Phase Space Geometry and Chaotic Attractors in Dissipative Nambu Mechanics}
\author{Zacharias Roupas}
\ead{roupas@inp.demokritos.gr}
\address{Institute of Nuclear Physics, N.C.S.R. Demokritos, GR-15310 Athens, Greece} 
\address{and}
\address{Physics Department, National Technical University of Athens, GR-15780, Athens, Greece}

\begin{abstract}
Following the Nambu mechanics framework we demonstrate that the non dissipative part of the Lorenz system can be generated 
by the intersection of two quadratic surfaces that form a doublet under the group $SL(2,\mathbb{R})$. 
All manifolds are classified into four distinct classes; parabolic, elliptical, cylindrical and hyperbolic. 
The 
Lorenz attractor is localized by a specific infinite set of one parameter family of these surfaces. The different classes correspond to different physical systems. The Lorenz system is identified as a charged rigid body in a uniform magnetic field with external torque and this system is generalized to give new strange attractors. 
\end{abstract}
%\noindent{\it Keywords: Lorenz attractor, Nambu mechanics, rigid body} 

\maketitle

\section{Introduction}\label{sec:intro}

We discuss the geometric structure of the integrable part of a chaotic three dimensional dynamical system. Our approach is generic, however we concentrate on the Lorenz model \cite{lor}. The analysis is performed in the Nambu mechanics \cite{nam} formalism. We find that important information regarding the boundary of the strange attractor can be extracted from this geometric structure. In addition, the symmetry that underlies the phase space geometry leads to mappings between different dynamical systems. Let us review some basic concepts. \\
\indent In \cite{nam} Y. Nambu introduced a formalism of classical mechanics to three dimensional phase spaces. In his words ``the Liouville theorem is taken as the guiding principle". Hence, the formalism applies to systems that preserve the phase space volume, i.e. $\nabla \vec{\upsilon} = 0$. The motion of points  $\vec{x}=(x,y,z) \in \mathbb{R}^3$ in phase space is determined by two functions of $(x,y,z)$, called Nambu Hamiltonians $H_1$ and $H_2$, by the equations:
\begin{equation}\label{nam:flow}
	\dot{\vec{x}} = \nabla H_1 \times \nabla H_2
\end{equation}
By definition \eref{nam:flow} the flow is volume preserving, since $\nabla \vec{\upsilon} = 0$, 
$\vec{\upsilon} = \dot{\vec{x}}$.
In general any function $F(x,y,z)$ is evolving according to
\begin{equation}\label{nam:F}
	\dot{F} = \nabla F\cdot (\nabla H_1 \times \nabla H_2) = \varepsilon_{ijk} \partial_i F \partial_j H_1 \partial_k H_2
\end{equation}
where $\varepsilon_{ijk}$ is the Levi-Civita tensor. The right-hand side defines the Nambu three bracket:
\begin{equation}\label{nam:nambuBr}
	\{ F,H_1,H_2\} = \varepsilon_{ijk} \partial_i F \partial_j H_1 \partial_k H_2
\end{equation}
The Nambu Hamiltonians are conserved quantities as one can see from \eref{nam:F}. The phase space orbit can therefore be realized as the intersection of the surfaces $H_1 = H_1(t=0)$ and $H_2 = H_2(t=0)$. \\
\indent In \cite{nam:tak, Axenides:2008rn} the Nambu bracket is realized as the Poisson bracket, denoted as $\{ \cdot ,\cdot \}_{H_2}$, induced on the manifold $H_2 = H_2(t=0)$, which is now upgraded to a phase space. We will call this bracket Nambu-Poisson bracket and is defined as in \eref{nam:nambuBr}. It is antisymmetric and satisfies the Jacobi identity. The Nambu-Poisson bracket of one dynamical variable $x_i$ and $H_1$ is:
\[
	\{ x_i ,H_1 \}_{H_2} = \varepsilon_{qjk} \partial_q x_i \partial_j H_1 \partial_k H_2 =
			\varepsilon_{ijk} \partial_j H_1 \partial_k H_2 \equiv \nabla H_1 \times \nabla H_2
\]
The dynamical variables form an algebra on $H_2$
\begin{equation}\label{nam:br}
	\{ x_i,x_j\}_{H_2} = \varepsilon_{qpk} \partial_q x_i \partial_p x_j \partial_k H_2 = \varepsilon_{ijk} \partial_k H_2
\end{equation}
The evolution of the system is given by
\begin{equation}\label{nam:eom2}
	\dot{x}_i = \{ x_i,H_1\}_{H_2}
\end{equation}
The surface $H_2(x,y,z) = const. = H_2(t=0)$ can in some cases be regarded as the group 
		manifold of a Lie group, for example if $H_2 = \frac{1}{2}x^2 + \frac{1}{2}y^2 + \frac{1}{2}z^2$. In this specific example $H_2$ is just the Casimir of $SO(3)$
		, so it is reasonable to call it $C$ and we will do so often in this work, calling $H_1$ as just $H$ as well (in some cases we will stick to $H_1$, $H_2$ notation). Now, $C$ defines an algebra of generalized coordinates $x,y,z$ by equation \eref{nam:br}:
		\[
		\begin{array}{l}
	\{x,y\}_{C} = z \\

	\{y,z\}_{C} = x \\

	\{z,x\}_{C} = y 
\end{array}
		\]
The evolution lies on surface $C = C(t=0) = const.$, with dynamics determined by $H$, or equivalently one can think of the
		operator $\{\bullet, H\}_{C}$ as generating the flow by repeated operations of a group element of the group with Casimir $C$ (which group element is determined by $H$). Therefore, in this case, the surface on which the system is constrained is a group manifold. 
		
		There is a freedom \cite{nam} in determining $H_1$, $H_2$ since any transformation $(H_1',H_2') = (H_1'(H_1,H_2),H_2'(H_1,H_2))$ with Jacobian determinant equal to one leaves the equations of motion invariant: 
\begin{eqnarray}\label{intro:sl2r}
\fl	\nabla H'_1 \times \nabla H'_2 = \varepsilon_{ijk} \partial_j H'_1 \partial_k H'_2 \nonumber\\  
		\lo= \varepsilon_{ijk} \left( \frac{\partial H'_1}{\partial H_1} \partial_j H_1 +  
	\frac{\partial H'_1}{\partial H_2} \partial_j H_2 \right) 
		\left( \frac{\partial H'_2}{\partial H_1} \partial_k H_1 +  
		\frac{\partial H'_2}{\partial H_2} \partial_k H_2\right) \nonumber\\
\lo=		 \varepsilon_{ijk}\left( \frac{\partial H'_1}{\partial H_1} \frac{\partial H'_2}{\partial H_2} -  
	\frac{\partial H'_1}{\partial H_2} \frac{\partial H'_2}{\partial H_1} \right)\partial_j H_1 \partial_k H_2 \nonumber\\
\lo=  |\frac{\partial(H'_1,H'_2)}{\partial(H_1,H_2)}|  \nabla H_1 \times \nabla H_2
\end{eqnarray}
We will call this freedom a gauge freedom. In the case of linear transformations the gauge group is $SL(2,\mathbb{R})$ and therefore the Nambu Hamiltonians form a doublet 
$
	h = (H_1 , H_2)
$
that transforms under $SL(2,\mathbb{R})$. 

In this work we will use a decomposition of the Lorenz system \eref{lor}, which has three phase space dimensions, to a conservative (non-dissipative) part and a dissipative part \cite{lor:nb}. 
In Hamiltonian mechanics (even phase space dimensions) it is common to add to some known Hamiltonian physical system, forced and dissipative terms. However, it is rarely the case that given the 
equations of motion of a forced, dissipative system ($\nabla \vec{\vv} \neq 0$), one is asked to guess a decomposition in a Hamiltonian (non-dissipative) part and a forced, dissipative part. This decomposition is of course not unique, but in Hamiltonian mechanics it is often
simple and natural to guess `the most physical' decomposition. For example, suppose one is given the system
	\begin{equation*}
		\begin{array}{lll}
    	    \dot{x}_1 = x_2 \\
        	\dot{x}_2 = -\omega x_1 - b x_2
		\end{array}	
	\end{equation*}
	which is a dissipative system ($\nabla\vec{\upsilon} = -b$). It may be decomposed to a non-dissipative part 
	$\vec{\upsilon}_{ND}$ with $\nabla \vec{\vv}_{ND} = 0$ and a dissipative part $\vec{\upsilon}_D$: 
	\[
\vec{\upsilon} = \vec{\upsilon}_{ND} + \vec{\upsilon}_{D}	
	\]
	 One may choose the parts $\vec{\upsilon}_{ND} = (0,-\omega x_1)$ and $\vec{\upsilon}_D = (x_2,-b x_2)$
	or even $\vec{\upsilon}_{ND} = (x_2 + x_2^2,-\omega x_1)$ and $\vec{\upsilon}_D = (-x_2^2,-b x_2)$. Mathematically they are nice equivalent choices but as far as physics is concerned `the most physical' choice would have been $\vec{\upsilon}_{ND} = (x_2,-\omega x_1)$ to get the harmonic oscillator for the non-dissipative part and to get the friction force $F = -b\times (velocity)$ for the remaining dissipative part $\vec{\upsilon}_D = (0,-b x_2)$. 
	
	The situation is analogous in three phase space dimensions using Nambu mechanics, even though one is not guided by intuition to find a physical decomposition. The first time a decomposition in Nambu mechanics was applied was in \cite{lor:nb} (to the Lorenz system). In \cite{Axenides:2009rh} has been set a general framework of the dissipative Nambu mechanics with more examples. The decomposition is chosen in \cite{Axenides:2009rh} so as $\nabla\vec{\upsilon}_{ND} = 0$ and ${\nabla}\times \vec{\upsilon}_{D} = 0$. It is not unique, however one is restricted by convenience according to motivations. The velocity field can always, at least locally, 
	(see \cite{Axenides:2009rh} and references therein) be written as
	\[
			\vec{\upsilon} = {\nabla}\times \vec{A} + {\nabla} D \quad , \mbox{ with}\quad \vec{\upsilon}_{ND} = {\nabla}\times \vec{A}
			\quad \mbox{and}\quad \vec{\upsilon}_{D} ={\nabla} D
	\]
	for some vector field $\vec{A}$ and some scalar field $D$. The field $\vec{A}$ has a gauge degree of freedom $\alpha(\vec{x})$
	\[
		\vec{A} \rightarrow \vec{A} + {\nabla}\alpha
	\]
	Any vector field can be written $\vec{A} = {\nabla}\alpha + \beta {\nabla}\gamma$ \cite{aris} for some scalar
	functions $\alpha$, $\beta$, $\gamma$. Working with specific Hamiltonians $H_1$ and $H_2$ means that you have chosen the gauge $\vec{A} = H_1 {\nabla}H_2$ for $\beta = H_1$ and $\gamma = H_2$. This gauge is known as the `Clebsch-Monge' gauge and the scalar functions $\alpha$, $\beta$, $\gamma$ are usually called `Monge's potentials' (see also \cite{Axenides:2011jc,Floratos:2011ct} for more details on the application of Nambu mechanics on non-Hamiltonian chaos and the quantum, non-commutative approach).
	
	\begin{figure}[b!]
\begin{center}
	\subfigure[The phase space trajectory of a damped harmonic oscillator. With red the non-dissipative trajectory (harmonic oscillator) corresponding to the initial conditions.]{ \label{fig:a}\includegraphics[width=55mm,height=40mm]{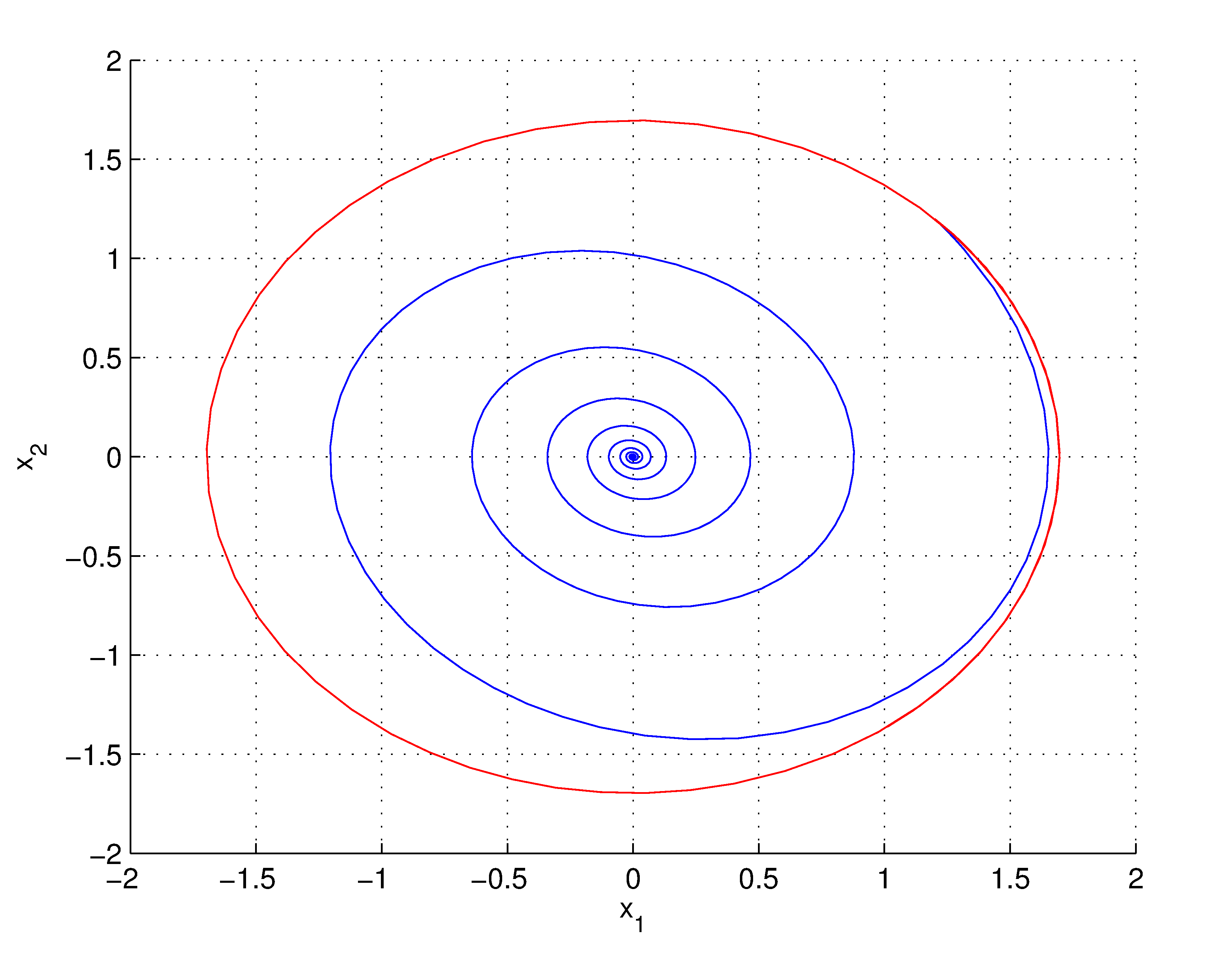} } 
\quad 
	\subfigure[The Lorenz system for large initial $(X_1,X_2,X_3)$ projected on $X_1-X_2$ plane. The spiral 
	is evolving towards decreasing $X_3$. With red the non-dissipative trajectory corresponding to the initial conditions.]{ \label{fig:b}\includegraphics[width=55mm,height=40mm]{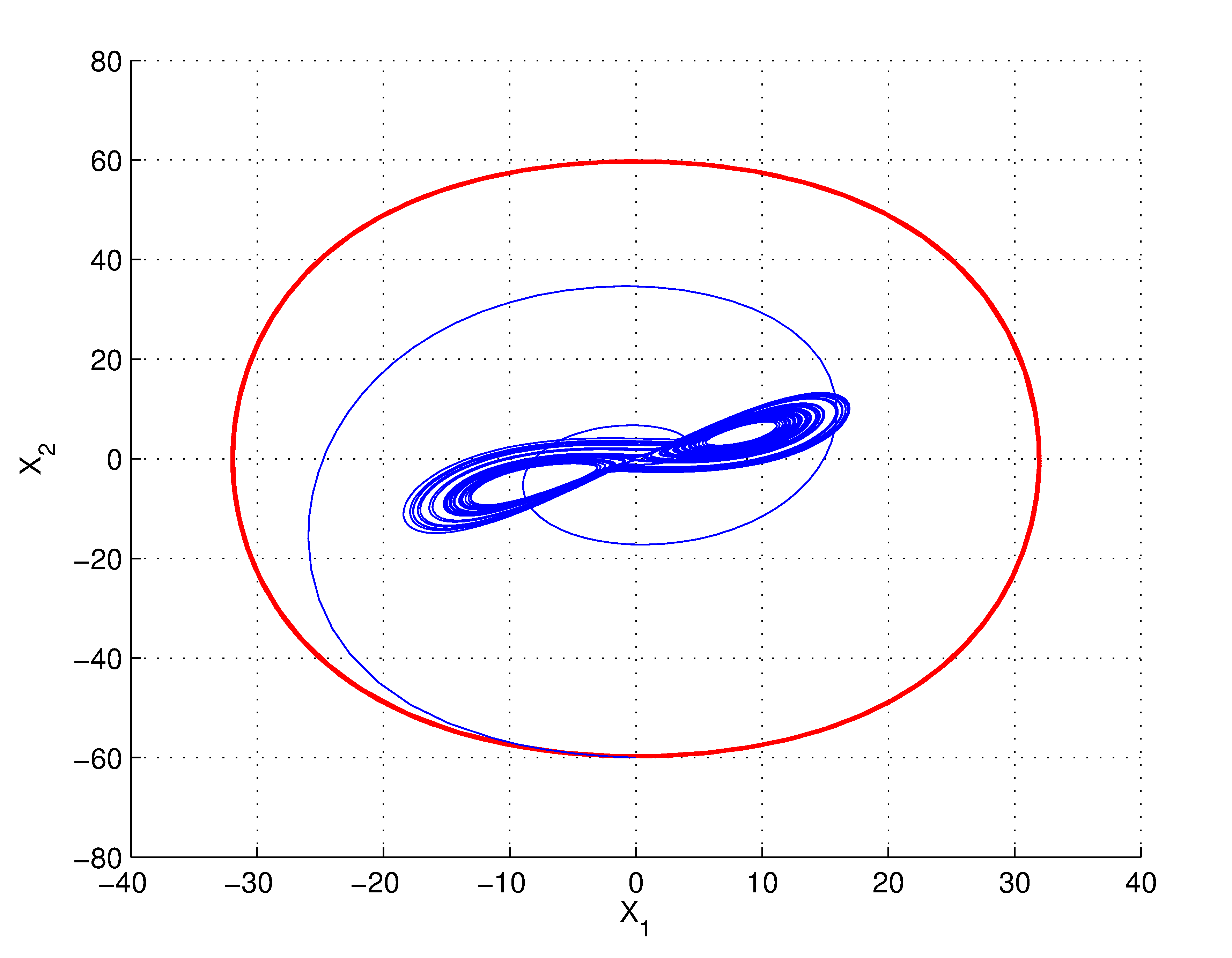} }
\caption{\textit{The damped harmonic oscillator and the Lorenz system. In the Lorenz system, despite of being a dissipative system ($\nabla \vec{\vv} < 0$), there is a constant driven term that appears in specific coordinates. For large $X_i$'s, however, the Lorenz system behaves as a purely dissipative (not driven) system. } 
\label{fig:hol}}
\end{center}
\end{figure}
	
	We are interested, and will focus in this work,	in the gauge freedom $SL(2,\mathbb{R})$ of $H_1$ and $H_2$.
	The application is on the Lorenz system, due to its unique 
	significance regarding chaotic dynamics and the interesting physical properties of its non-dissipative part in the specific decomposition we will use.  We classify the `Nambu doublets',
	i.e. the pair of Hamiltonians, of the Lorenz non-dissipative part with criterion the isomorphism of the algebras. We find four different classes of doublets in section \ref{sec:class} that are identified with four different
	systems in section \ref{sec:physical}, which are therefore dynamically equivalent. Three of them are very familiar physical systems, namely the simple pendulum, the Duffing oscillator and a charged dielectric rigid body inside 
	a homogeneous magnetic field, with the fourth being an $SO(2,1)$ system.  
	We may say that these four systems are `dual' to each other. 
	
	We stress that Lorenz defines his system as a forced, dissipative system \cite{lor}. He starts from an abstract, conservative system with some conservative quantity $Q$ which can be a compact `\textit{shell like}' surface. Then he makes the hypothesis that `\textit{[...], whenever $Q$ equals
		or exceeds some fixed value $Q_1$, the dissipation acts to diminish $Q$ more rapidly then the forcing can increase $Q$, then
		$(-dQ/dt)$ has a positive lower bound where $Q \geq Q_1$, and each trajectory must ultimately become trapped in the region where $Q<Q_1$.}' quoted
		straight from the famous 1963 paper \cite{lor}. This reasoning implies that the conserved quantities of the corresponding
		conservative system, which in our decomposition are the Nambu functions\footnote{We sometimes call the Nambu Hamiltonians of the non-dissipative system, just `Nambu functions'.}, is most natural to be used for finding trapping surfaces and for the localization of the attractor in general. This is what we do in section \ref{sec:local} using the Nambu functions of section \ref{sec:class}. 
		
In section \ref{sec:so3} we find that the Lorenz system is equivalent to a physical system that presents this type of behaviour, quoted earlier, for the forced, damped terms. This system is the $SO(3)$ system we mentioned (charged rigid body in a magnetic field), but now with external torques. The generalized coordinates are the three components of the angular momentum $\vec{L}$.  
	The conservative function ``$Q$" of Lorenz
		is just	the measure of angular momentum $Q = |\vec{L}|$ and its conservation for the conservative part expresses the conservation of 
		angular momentum. The damped terms are torques proportional to angular momentum, like air resistance. There is a constant torque along $Z$-direction that acts as a driven(forced) term when $L_z<0$. For large $\vec{L}$ the 
		friction terms dominate over the constant torque (see \Fref{fig:hol}).
		The attractor in this coordinates lies in $L_z < 0$. 
	Since there	exists a constant driven term for $L_z$, the $L_z$ angular momentum could not ever become constantly zero.

	This system \eref{sph:lor} is generalized for arbitrary direction of the magnetic field to give a new dynamical system. This general system \eref{sph:rsys} presents
	new strange attractors given in Figures \ref{fig:rat1}, \ref{fig:rat3}.

\section{Lorenz non dissipative part and $SL(2,\mathbb{R})$ classification}\label{sec:class}
	
\begin{figure}[tb!]
\begin{center}
\subfigure[]{\label{fig:interval}\includegraphics[width=40mm,height=30mm]{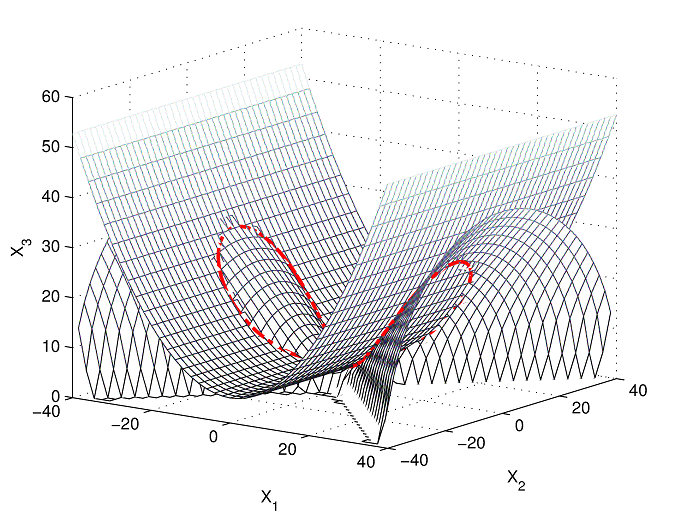}} 
\quad 
\subfigure[]{\label{fig:inter_b}\includegraphics[width=40mm,height=30mm]{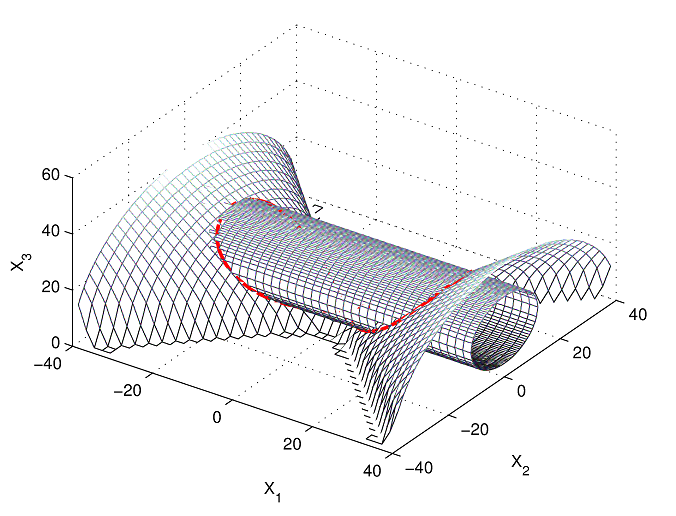}}
\quad
\subfigure[]{\label{fig:inter_c}\includegraphics[width=40mm,height=30mm]{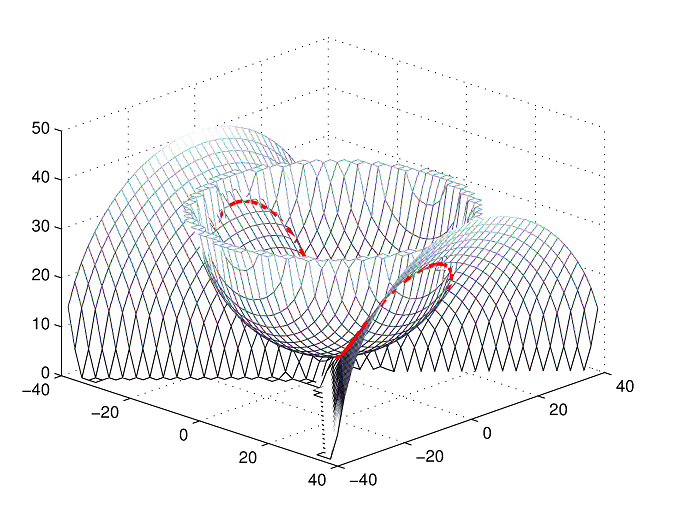}} 
\\
\subfigure[]{\label{fig:inter_d}\includegraphics[width=40mm,height=30mm]{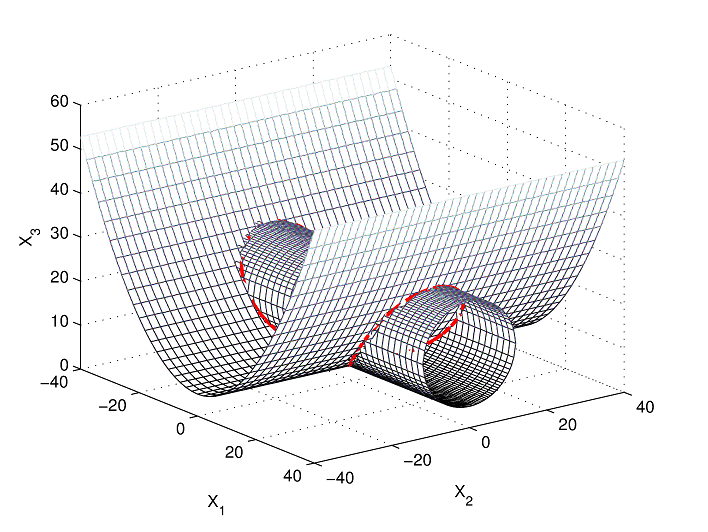}} 
\quad 
\subfigure[]{\label{fig:inter_e}\includegraphics[width=40mm,height=30mm]{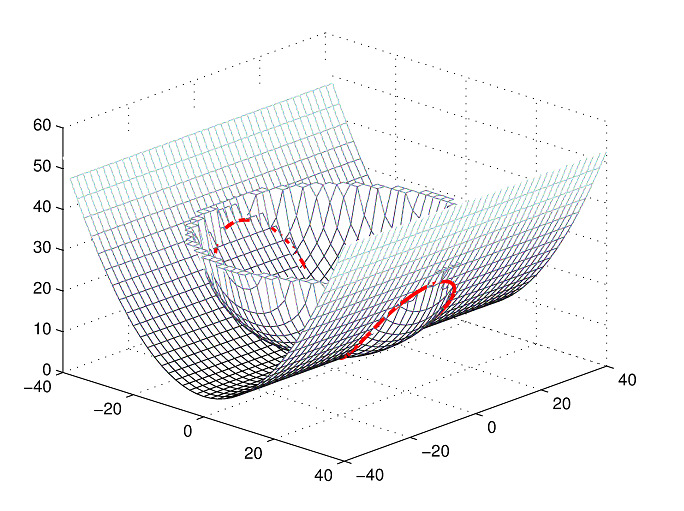}}
\quad
\subfigure[]{\label{fig:inter_f}\includegraphics[width=40mm,height=30mm]{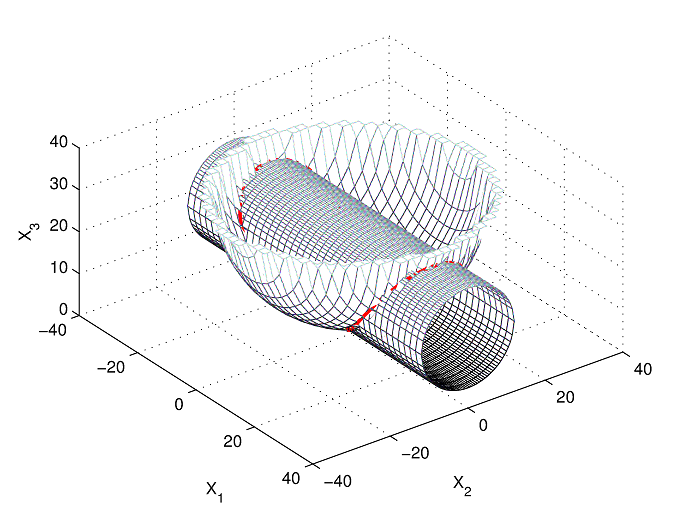}} 
\caption{An orbit, in red, of the conservative system \eref{consLor} as defined by the intersection of various Nambu surfaces (same orbit at each case).\label{fig:inter}}
\end{center}
\end{figure}

The Lorenz system \cite{lor}, defined by:
\begin{equation}\label{lor}
     \begin{array}{ll}
        \dot{X} =  \sigma Y 	&-\sigma X \\
        \dot{Y} = - XZ 	+ rX	&- Y \\
		\dot{Z} =  XY 		&-b Z
		\end{array}
\end{equation}
where $r$, $\sigma$ and $b$ are positive constants\footnote{
In every figure of the Lorenz system the  free parameters are assumed to be $r=28$, $\sigma = 10$ and $b=8/3$, unless stated differently.} 
is a dissipative system, since $\nabla \vec{\vv} = -(\sigma + 1 + b) < 0$ where $\vec{\vv} = (\dot{X},\dot{Y},\dot{Z})$. In \cite{lor:nb} the following decomposition is proposed:
\begin{equation}
	\vec{\vv}_{ND} = (\sigma Y, -XZ +rX,XY) \; , \; \vec{\vv}_{D} = (-\sigma X, -Y,-bZ)
\end{equation}
with the full system given by $\vec{\vv}=\vec{\vv}_{ND} + \vec{\vv}_D$. In the same work two Nambu Hamiltonians for the non-dissipative part $\vec{\vv}_{ND}$ are found 
\begin{equation}\label{cla:ham1}
	H = \frac{1}{2}Y^2 + \frac{1}{2}Z^2 - r Z
\; , \;
	C = -\frac{1}{2}X^2 + \sigma Z
\end{equation}
 by integration of the equations:
 	\begin{equation}
		\frac{dY}{dZ} = \frac{r-Z}{Y} \quad,\quad \frac{dZ}{dX} = \frac{X}{\sigma} 
	\end{equation}		
To decouple the parameter dependence between the two parts (non-dissipative and dissipative) and for reasons of convenience for our purposes we introduce the scaling transformation:
\begin{equation}\label{resc}
%\begin{array}{c}
		X = X_1 \; , \;
		Y = \sqrt{r/\sigma} X_2 \; , \; 
		Z = \sqrt{r/\sigma} X_3 \; , \;
		\rho = \sqrt{\sigma r}
%\end{array}
\end{equation}
and the Lorenz system \eref{lor} becomes:
\begin{equation}\label{fullLor}
\begin{array}{ll}
    \dot{X}_1 = \rho X_2 &- \sigma X_1 \\
    \dot{X}_2 =     -X_1X_3 + \rho X_1 &- X_2 \\
	\dot{X}_3 =	 X_1 X_2  &- bX_3
\end{array}
\end{equation}
We will work independently on the non dissipative part:
\begin{equation}\label{consLor}
\begin{array}{l}
    \dot{X}_1 = \rho X_2 \\
    \dot{X}_2 =     -X_1X_3 + \rho X_1  \\
	\dot{X}_3 =	 X_1 X_2  
\end{array}
\end{equation}
The Nambu Hamiltonians \eref{cla:ham1} become in the new rescaled variables:
\begin{equation}\label{class:cylpar}
		H =  \frac{1}{2} X_2^2 + \frac{1}{2} X_3^2 - \rho X_3 \; , \;
		C = -\frac{1}{2} X_1^2 + \rho  X_3 
\end{equation}
We will call this pair of functions a `Nambu doublet'
\begin{equation}
	h = \left(
		\begin{array}{c}
		H \\
		C
		\end{array}
	\right)
\end{equation}
since it transforms as a whole under a restricted class of transformations. These are all transformations 
with Jacobian determinant equal to one, as we saw in section \ref{sec:intro}. We will restrict ourselves in linear transformations so as to work only with quadratic functions.
In this case, the transformation group is the group of real $2\times 2$ matrices with determinant one, namely the
$SL(2,\mathbb{R})$. For example let generate a different doublet by action of an $SL(2,\mathbb{R})$ matrix:
\begin{equation}
	h' =
A
h \Rightarrow
	\left(
\begin{array}{c} 
	H' \\
	C'
\end{array}
\right)
=
	\left(
\begin{array}{cc} 
	0 & 1 \\
	-1 & 0
\end{array}
\right)
	\left(
\begin{array}{cc} 
	H \\
	C
\end{array}
\right)
=
	\left(
\begin{array}{cc} 
	C \\
	-H
\end{array}
\right)
\end{equation}
We see that the role of $H$ and $C$ can be interchanged, as long as you change the sign of one of the functions, since we could 
have used 
\[
A = 	\left(
\begin{array}{cc} 
	0 & -1 \\
	1 & 0
\end{array}
\right)
\]
as well\footnote{
It is apparent that the system is invariant in these cases since
$
\dot{X}_i = \nabla H \times \nabla C = (-\nabla C) \times \nabla H = \nabla C \times (-\nabla H)
$
}.
Another transformation could be:
\[
		\left(
\begin{array}{c} 
	H_0 \\
	C_0
\end{array}
\right)
=
	\left(
\begin{array}{cc} 
	0 & -1 \\
	1 &  1
\end{array}
\right)
	\left(
\begin{array}{cc} 
	H \\
	C
\end{array}
\right)
\]
that gives:
\begin{equation}\label{class:parhyp}
\left(
\begin{array}{c} 
	H_0 \\
	C_0
\end{array}
\right)
=
\left(
\begin{array}{c} 
	 \frac{1}{2} X_1^2 - \rho  X_3  \\
	 -\frac{1}{2} X_1^2 + \frac{1}{2} X_2^2 + \frac{1}{2} X_3^2
\end{array}
\right)
\end{equation}
Recall from the Introduction, that the Nambu Hamiltonians $H$, $C$ of a volume preserving system define surfaces in phase space
because they are conserved. Once initial conditions are given, the solution of the system should lie on both surfaces defined by
\begin{equation}
\left(
\begin{array}{c} 
	H(X_1,X_2,X_3) \\
	C(X_1,X_2,X_3)
\end{array}
\right)
=
\left(
\begin{array}{c} 
	H(X_1(0),X_2(0),X_3(0)) \\
	C(X_1(0),X_2(0),X_3(0))
\end{array}
\right)
\end{equation}
Therefore the intersection of the two Hamiltonians defines the phase space trajectory (see Figure \ref{fig:inter}). Thus, given a Nambu doublet and 
initial conditions, it is not only given the equations of motion (velocity field) but the solution of the system, as well!

In equation \eref{class:parhyp} is defined the Nambu Hamiltonian function $C_0$ that defines a hyperboloid, once
initial conditions are given. This is a topologically different surface from the cylinder and parabolic surface of equation
\eref{class:cylpar}. An $SL(2,\mathbb{R})$ transformation led to a new type of surface. This is reflected to the algebra
of the Nambu bracket, since:
\begin{equation}
	\begin{array}{l}
	\{X_1,X_2\}_C = \rho \\

	\{X_2,X_3\}_C = -X_1 \\

	\{X_3,X_1\}_C = 0 
\end{array}
\end{equation}
while:
\begin{equation}\label{class:so21}
	\begin{array}{l}
	\{X_1,X_2\}_{C_0} = X_3 \\

	\{X_2,X_3\}_{C_0} = -X_1 \\

	\{X_3,X_1\}_{C_0} = X_2 
\end{array}
\end{equation}
The later algebra is the $SO(2,1)$ algebra\footnote{
Using the algebra \eref{class:so21}, the properties of the Poisson bracket and $H_0$ one can show that the doublet \eref{class:parhyp} gives the system \eref{consLor}:
\[
	\begin{array}{l}
		\dot{X}_1=\{X_1,H_0\}_{C_0} = \{X_1,\frac{1}{2}X_1^2\}_{C_0} + \{X_1,-\rho X_3\}_{C_0} = \rho X_2 \\
		\dot{X}_2=\{X_2,H_0\}_{C_0} = \{X_2,\frac{1}{2}X_1^2\}_{C_0} + \{X_2,-\rho X_3\}_{C_0} = -X_1X_3 + \rho X_1 \\
		\dot{X}_3=\{X_3,H_0\}_{C_0} = \{X_3,\frac{1}{2}X_1^2\}_{C_0} + \{X_3,-\rho X_3\}_{C_0} = X_1X_2
	\end{array}
\]
}. We could get an isomorphic to $SO(2,1)$ algebra by applying to \eref{class:parhyp} transformations of the form
\begin{equation}
	A = 
	\left(
\begin{array}{cc} 
	\zeta & 0 \\
	0 &  1/\zeta
\end{array}
\right)	
\end{equation}
for some $\zeta \in \mathbb{R}$. These algebras would correspond to other hyperboloids:
\[
	C' = - \frac{1}{2}\frac{1}{\zeta}X_1^2 + \frac{1}{2} \zeta X_2^2 + \frac{1}{2}\zeta X_3^2 - (\zeta - \frac{1}{\zeta})\rho X_3
\]
One may ask what other types of quadratic surfaces one may get. We answer this question and classify the doublets
according to isomorphisms of the algebras.

Let an arbitrary $SL(2,\mathbb{R})$ matrix 
\[
A = 
\left(
\begin{array}{cc} 
	\alpha & \beta \\
	\gamma & \delta
\end{array}
\right)
\; , \;
\alpha \delta - \beta\gamma = 1
\]
act on the doublet \eref{class:parhyp}. Then we get a general expression for the possible quadratic doublets\footnote{
Note that there does not exist another different distinct set of quadratic doublets. If it existed would mean that members
of one set are not related with members of the other set by transformations with Jacobian determinant one, which is not allowed
as can be realized by \eref{intro:sl2r}. In addition  quadratic doublets can only be related with linear $SL(2,\mathbb{R})$ transformations with each other, because any non-linear transformation would raise the rank of the functions.
}:
\begin{equation}\label{doublet}
	h = 
\left(	\begin{array}{c} 
		S(X_1,X_2,X_3;\alpha ,\beta) \\
		S(X_1,X_2,X_3;\gamma ,\delta)
	\end{array}
\right)
\quad \forall \alpha,\beta,\gamma,\delta \in \mathbb{R} : \alpha \delta - \beta\gamma = 1
\end{equation}
with
\begin{equation}\label{sab}
	S(X_1,X_2,X_3;\alpha ,\beta) = (\alpha - \beta)\frac{1}{2}X_1^2 + \beta\frac{1}{2}X_2^2 + \beta\frac{1}{2}X_3^2 - \alpha \rho X_3
\end{equation}
Of course it does not matter with what doublet one may begin, since one can always with a re-parametrization end up with equation \eref{sab}. The expression for $S(X_1,X_2,X_3;\gamma ,\delta)$ is identical with \ref{sab} and
describes exactly the same surfaces. The $S(\alpha ,\beta)$ and $S(\gamma ,\delta)$ are different only when
considered as part of a doublet, because it is then imposed the restriction $\alpha \delta - \beta\gamma = 1$.

\begin{figure}[t!]
\begin{center}
	\subfigure[Parabolic]{\label{fig:geoP}\includegraphics[width=50mm,height=40mm]{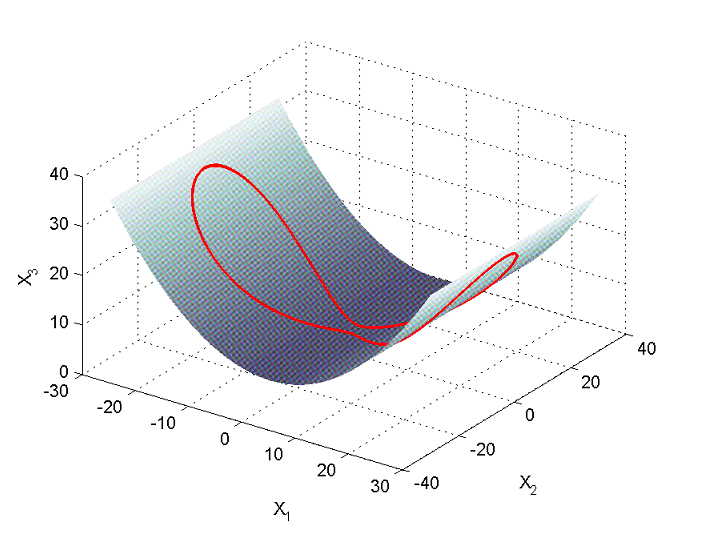}} 
\quad 
	\subfigure[Hyperbolic]{\label{fig:geoH}\includegraphics[width=50mm,height=40mm]{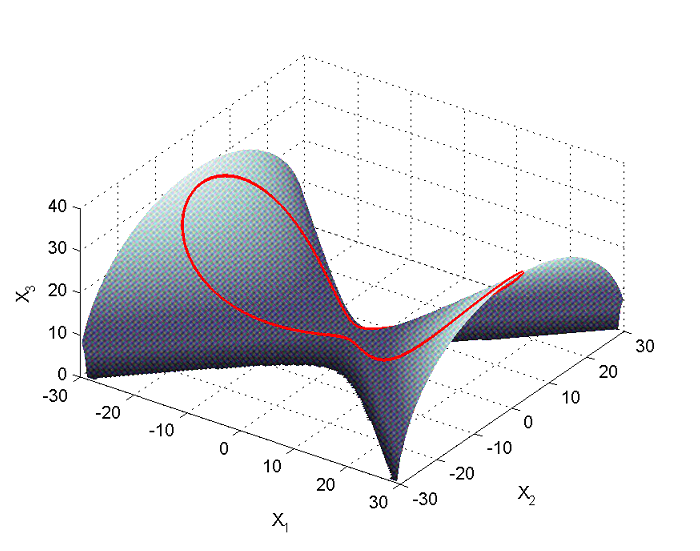}}
	\\
	\subfigure[Cylindrical]{\label{fig:geoC}\includegraphics[width=50mm,height=40mm]{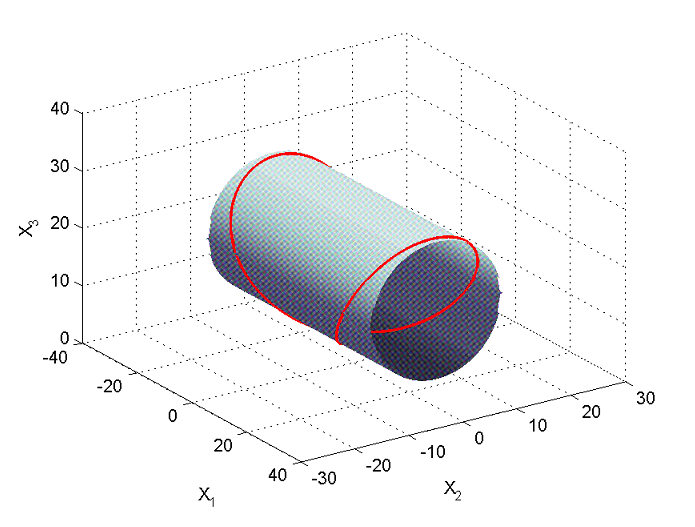}} 
\quad 
	\subfigure[Elliptical]{\label{fig:geoS}\includegraphics[width=50mm,height=40mm]{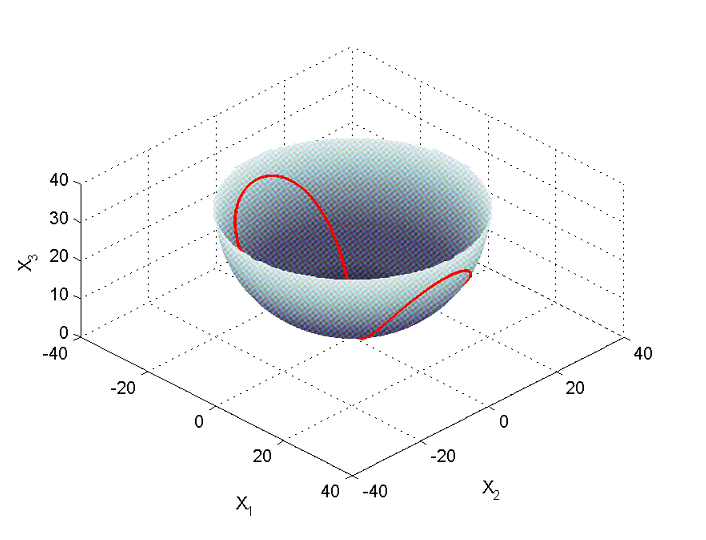}}
\caption{The four different classes of the Nambu surfaces (invariant manifolds) of the non dissipative part \eref{consLor} of the Lorenz system. An orbit, in red, of the system lies on any of these surfaces. 
\label{fig:geo}}
\end{center}
\end{figure} 

\begin{table}[htbp]
\begin{center}
\begin{small}
\begin{tabular}{ll}
\hline\noalign{\smallskip}
\multicolumn{2}{c}{$C_{(\lambda)} = (\lambda - 1)\frac{1}{2}X_1^2 + \frac{1}{2}X_2^2 + \frac{1}{2}(X_3 - \lambda \rho )^2$}
\\[1.2ex]
$\lambda < 1$ & : a set of hyperboloids centred at $X_3^c = \lambda \rho$ \\[1.2ex]
$\lambda > 1$ & :  a set of ellipsoids centred at $X_3^c = \lambda \rho$ \\[1.2ex]
$\lambda = 1$ & : a cylinder centred at $X_3^c = \rho$ \\[1.2ex]
$\lambda \rightarrow \pm \infty$ & : a parabolic cylinder \\[1.2ex]
\hline
\end{tabular}
\end{small}
\caption{The Nambu surfaces}
\label{tab:notation}
\end{center}
\end{table}
\indent The possible types of quadratic surfaces are given by equation \eref{sab}. These determine the possible Nambu algebras and the geometry of the phase-space through the Nambu bracket:
\[
	\{ X_i,X_j\}_S = \epsilon_{ijk}\partial_k S
\]
Let $\lambda = \frac{\alpha}{\beta}\in \mathbb{R}$ and
\begin{equation}\label{casimirs}
	C_{(\lambda)} = (\lambda - 1)\frac{1}{2}X_1^2 + \frac{1}{2}X_2^2 
			+ \frac{1}{2}(X_3 - \lambda \rho )^2
\end{equation}
where $C_{(\lambda)}$ is an abbreviation for $C(X_1,X_2,X_3;\lambda)$.
Then, up to a constant shift it is:
\[
	S(X_1,X_2,X_3;\alpha ,\beta) = \beta C_{(\lambda)}
\]
Rescaling of a Nambu function: 
$
	H_1' = \beta H_1
$
leaves the corresponding Nambu surface invariant since:
$
	H_1' = H_1'(0) \Leftrightarrow \beta H_1 = \beta H_1(0) \Leftrightarrow H_1 = H_1(0)
$
so that the equations $H_1' = H_1'(0)$ and $H_1 = H_1(0)$ define exactly the same surface in the phase space. Therefore equation \eref{casimirs} describes all possible quadratic surfaces, which are listed in \Tref{tab:notation} and can be seen in \Fref{fig:geo}. From the algebraic point of view the rescaling leads to isomorphic algebras. We see that there are infinite surfaces, of four types, when classified with respect to 
the isomorphisms of the corresponding algebras. The surface given by $S(\alpha, 0)$ (that is $C$ of equation \eref{class:cylpar}) corresponds to $\lambda \rightarrow \pm \infty$.  
\begin{table}[tb!] %here
\footnotesize
	\centering	
\begin{tabular}{c}
\hline\noalign{\smallskip}
$
\begin{array}{c}
H = (\alpha - 1) \frac{1}{2}X_1^2 + \frac{1}{2}X_2^2 + \frac{1}{2}X_3^2 - \alpha \rho X_3 \; ,\; \forall \alpha\in \mathbb{R}
\\
\begin{array}{l}
	\{X_1,X_2\}_C = \rho \\

	\{X_2,X_3\}_C = -X_1 \\

	\{X_3,X_1\}_C = 0 
\end{array}
\end{array}
$
\\ 
\\
$
\begin{array}{c}
H = (1 - \beta) \frac{1}{2}X_1^2 + \beta\frac{1}{2}X_2^2 + \beta\frac{1}{2}X_3^2 - \rho X_3\; , \;\forall \beta\in \mathbb{R}
\\
\begin{array}{cc}
SO(2,1)
&
\begin{array}{l}
	\{X_1,X_2\}_{C_0} = X_3 \\

	\{X_2,X_3\}_{C_0} = -X_1 \\

	\{X_3,X_1\}_{C_0} = X_2 
\end{array}
\end{array}
\end{array}
$
\\ 
\\
$
\begin{array}{c}
H = \frac{1}{2}X_1^2 + \beta\frac{1}{2}X_2^2 + \beta\frac{1}{2}X_3^2 - \rho (\beta +1 )X_3 \; , \;\forall \beta\in \mathbb{R}
\\
\begin{array}{cc}
SE(2)
&
\begin{array}{l}
	\{X_1,X_2\}_{C_1} = X_3 - \rho \\

	\{X_2,X_3\}_{C_1} = 0 \\

	\{X_3,X_1\}_{C_1} = X_2 
\end{array}
\end{array}
\end{array}
$
\\ 
\\
$
\begin{array}{c}
H = (\beta + 1)\frac{1}{2}X_1^2 + \beta\frac{1}{2}X_2^2 + \beta\frac{1}{2}X_3^2 - \rho (2\beta +1 )X_3 \; , \;\forall \beta\in \mathbb{R}
\\
\begin{array}{cc}
SO(3)
&
\begin{array}{l}
	\{X_1,X_2\}_{C_2} = X_3 - 2\rho \\

	\{X_2,X_3\}_{C_2} = X_1 \\

	\{X_3,X_1\}_{C_2} = X_2 
\end{array}
\end{array}
\end{array}
$
\\[1.5ex]
\hline
\end{tabular}
\caption{ The Nambu doublets Classification}
\label{tab:doublets}
\end{table}
\\
\indent Let list the possible algebras. The hyperboloid centred at the origin $C_0$ gives the $SO(2,1)$ algebra
and any other hyperboloid gives an algebra isomorphic to $SO(2,1)$. \\
\indent The sphere $C_2 = \frac{1}{2}X_1^2 + \frac{1}{2}X_2^2 
			+ \frac{1}{2}(X_3 - 2 \rho )^2$ centred at $X_3^c = 2\rho$, after the transformation $X_3 \rightarrow X_3 - 2\rho$, gives the $SO(3)$ algebra
$
	\{X_i,X_j\}_{C_2} = \varepsilon_{ijk} X_k \; ,\;
$
and any ellipsoid, an algebra isomorphic to $SO(3)$. \\
\indent The cylinder $C_1 = \frac{1}{2}X_2^2 
			+ \frac{1}{2}(X_3 - \rho )^2$ gives the $SE(2)$ algebra after the transformation $X_3 \rightarrow X_3 - \rho$.
\\
\indent The algebra obtained by the parabolic cylinder $C_{(\infty)}$ call it $C = -\frac{1}{2}X_1^2 + \rho X_3$
can be realized as the Heisenberg algebra $\{X_1,X_2 \}_C = \rho$ of two independent variables $X_1,X_2$ on the curved plane $C$:
$
	\{X_2,X_3\}_{C} = -X_1 \; ,\;
	\{X_3,X_1\}_{C} = 0
$. \\
\indent With these four types we classify the possible Nambu doublets as in \Tref{tab:doublets} and an orbit defined by six different doublets can be seen in \Fref{fig:inter}. In \Tref{tab:doublets} one of the four possible types of 
algebras is chosen with a specific $C$ at each case and the corresponding $H$ is found by the condition $\alpha \delta - \beta\gamma = 1$.

\section{Localization of the Lorenz Attractor}\label{sec:local}

When the dissipative part $\vv_D = (-\sigma X_1,-X_2,-b X_3)$ is added to the system \eref{consLor}, the Lorenz attractor is formed. This is a global attractor, that is this subset of phase space that every trajectory of any initial conditions approaches for $t\rightarrow \infty$. By definition the attractor should lie in a bounded region and determining this region is what we call localization. The possible geometries of the boundary of the Lorenz attractor have been discussed by Lorenz \cite{lor,lor:loc1}, Sparrow \cite{lor:spa} and others \cite{lor:loc2,lor:loc3}. We will not find new surfaces, not mentioned 
in bibliography already. Our scope is only to prove that the Nambu functions found in previous section, i.e. the 
quadratic invariant manifolds of the non-dissipative part, define proper localization surfaces. 

Let us change the notation and use just $S$ for $2C_{(\lambda)}$ of equation \eref{casimirs}.
\begin{equation}\label{loc:S}
	S(X_1,X_2,X_3) = (\lambda - 1)X_1^2 + X_2^2 + (X_3 - \lambda \rho )^2 
\end{equation}
When $X_1$, $X_2$, $X_3$ are subject to the Lorenz evolution (with the dissipative part added), $S$ is not conserved but is changing with time:
	\begin{equation}
		\dot{S}	 = \frac{1}{2}b (\lambda\rho)^2  - 2\left(  \sigma (\lambda - 1)X_1^2 + X_2^2 
					+ b (X_3-\frac{1}{2}\lambda\rho)^2 \right)
	\end{equation}
	Assume for the moment $\lambda > 1$. Suppose there exists some real constant $K_{max}$ for which $\dot{S}$ can be positive only inside the ellipsoid $S = K_{max}$ and nowhere else (see Appendix C of Sparrow \cite{lor:spa} and Doering and Gibbon \cite{lor:loc2}, as well).
	 Then for any point outside this ellipsoid, $\dot{S} < 0$ and therefore $S$ will be decreasing until it gets $S < K_{max}$ i.e. the trajectory gets inside the ellipsoid. Of course, if the trajectory starts inside
	this ellipsoid, it will remain there, since on the outside is everywhere $\dot{S} < 0$. If such $K_{max}$ exists, it is the solution of the constrained maximization problem:
	\begin{equation}\label{loc:max}
		K_{max} = max\{S(X_1,X_2,X_3):\dot{S} \geq 0\} \quad ,\,\lambda \geq 1
	\end{equation}
		This is a typical problem to solve with the Lagrange multipliers method (\ref{app:lag}). This condition
		defines a compact attracting surface (in fact every other ellipsoid $S = K$ with
		$K>K_{max}$ is an attracting surface, since $\dot{S} < 0$ everywhere on the outside of all of these). The solution of 
		\eref{loc:max} is given in \eref{local} where $K_{max} = R_a^2$. 
		
\begin{figure}[tb!]
\begin{center}
	\includegraphics[width=70mm,height=50mm]{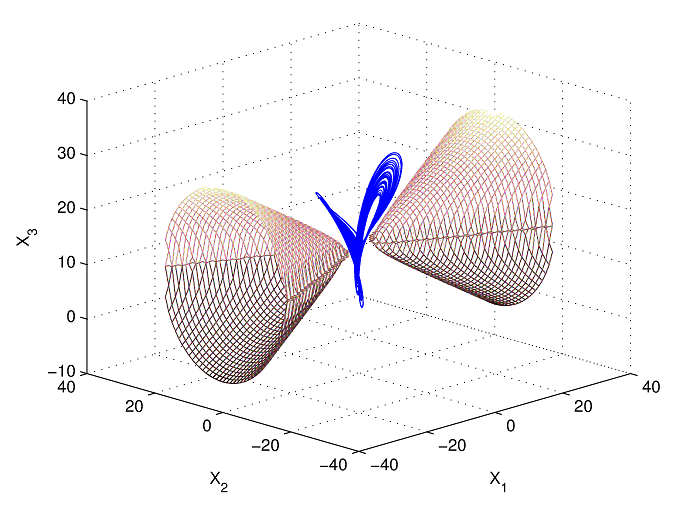} 
\end{center}
\caption{The repelling two-sheeted hyperboloid $\lambda = 0.8$ and the Lorenz attractor.}
\label{fig:repHypb}
\end{figure} 

We argue here that similar reasoning can be applied for non-positive-definite functions (non-compact surfaces). For $\lambda < 1$ the surfaces are hyperboloids which are non compact. In this case the equation $S=K$ for some constant $K$
defines a one sheeted hyperboloid if $K>0$ and a two sheeted hyperboloid if $K<0$. For simplicity, with no loss of
generality, assume that $K<0$. Then the condition $S>K$ defines the region outside the `lobes' of the two sheeted hyperboloid
(the attractor lies in this region in \Fref{fig:repHypb}), while $S<K$ defines the region inside the lobes. As previously, 
assume there exists some constant value $K_{min}$ such that $\dot{S}$ can be negative only outside the lobes (region $S>K_{min}$) and nowhere else. Then any trajectory starting from inside the lobes will be expelled to the outside, since $\dot{S}>0$ in the
inside and therefore $S$ is increasing until it gets $S>K_{min}$ i.e. the trajectory gets outside the lobes. Of course, if 
it starts outside the lobes, it is impossible to get inside. We call such a kind of surface, a repelling surface. The same reasoning can be applied to any non-compact surface. The problem is now a minimization problem\footnote{
The signs is a matter of convention, since we could have used a function $\tilde{S}=-S$ and turn the problem to a
maximization problem.  
}
		\begin{equation}\label{loc:min}
			K_{min} = min\{S(X_1,X_2,X_3):\dot{S} \leq 0\} \quad ,\,\lambda < 1
		\end{equation}
		The solution is given in \eref{local} where $K_{min} = -R_r^2$. In the \ref{app:lag} is given analytically the Lagrange's multipliers method. 
		
		Another way to understand localization 
		is to think of a fixed surface
		and imagine the flow at each point crossing it (Giacomini and Neukirch \cite{lor:loc3}). For a localization 
		surface the flow should cross every point
		of the surface towards the same direction (semi-permeable surface) \cite{lor:loc3}; inwards for the attracting (compact) surfaces and outwards for the repelling (non-compact) ones. Or else the trajectory could pass through the surface in one direction at some point	and get through the opposite direction at some other point of the surface. Thus, the scalar product of the vector normal 
		to the surface $\nabla{S}$ at each point and the vector tangent to the flow $\vec{\vv}$ should have a constant sign at each point. That is:
	\begin{equation}
		\dot{S} = (\nabla S)\cdot \dot{\vec{X}} 
	\end{equation}
	has constant sign at every point on a localization surface.
	This is proven in the \ref{app:nor} for all repelling surfaces of \Eref{local2}. 

\begin{figure}[t!]
\begin{center}
	\includegraphics[width=70mm,height=50mm]{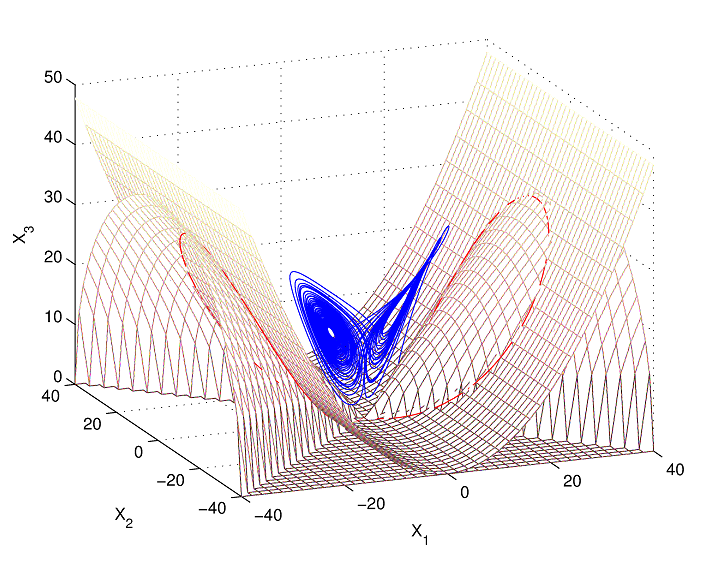}
\end{center}
\caption{The homoclinic orbits of the conservative system \eref{consLor} with red as defined by the repelling cone and parabolic cylinder. When the dissipative part is added, the flow is forced towards the Lorenz attractor.}
\label{fig:homA}
\end{figure}

Let summarize our results. We find the following attracting and repelling surfaces:
\begin{equation}\label{local}
\begin{array}{c}
	\fl S(X_1,X_2,X_3) = K \Leftrightarrow (\lambda - 1)X_1^2 + X_2^2 + (X_3 - \lambda \rho )^2 = K 
\\
	\fl\forall K \in \mathbb{R} \, : \,
	K
	\cases{
	\geq R_a^2 &for $\lambda \geq 1$  \\
	\leq \tilde{R}_r^2 &for $\{\lambda = 0,-\infty\}$ and $\{\lambda = (-\infty,0)$, $X_3>0\}$ \\ 
	\leq -R_r^2 &for $\lambda < 1$ } 
\\
	\fl \mbox{where: }\,
	R_a = 	
	\cases{
	\lambda \rho &for $b \leq 2$ \\
	\lambda \rho \frac{b}{2\sqrt{b-1}} &for $b >2$
	} 
	\, ,\,
	\tilde{R}_r = \lambda \rho 
	\, ,\,
	R_r = \lambda \rho\frac{b}{2\sqrt{\sigma (\sigma - b)}}
\end{array}
\end{equation}
The attracting cases hold for $\sigma > 1$, the repelling cases hold for $\sigma > 1$ and $\sigma > b$ except for the 
parabolic surface $\lambda \rightarrow -\infty$ that is $-X_1^2+2\rho X_3 = 0$ which holds for $\sigma > b/2$. The surfaces
that are closer to the attractor and provide the optimum localization are the extreme cases of \eref{local}
that correspond to the equalities:
\begin{equation}\label{local2}
	S = R_a^2 \; ,\; S = \tilde{R}_r^2 \; ,\; S = -R_r^2
\end{equation}
This means that the strange attractor lies in the region $\mathcal{A}$ that is defined as:
\begin{equation}
	\mathcal{A} = \{(X_1,X_2,X_3): S \leq R_a^2 \mbox{ and } S \geq \tilde{R}_r^2 \mbox{ and } S \geq -R_r^2 \}
\end{equation}

\begin{figure}[bt!]
\begin{center}
	\includegraphics[width=70mm,height=50mm]{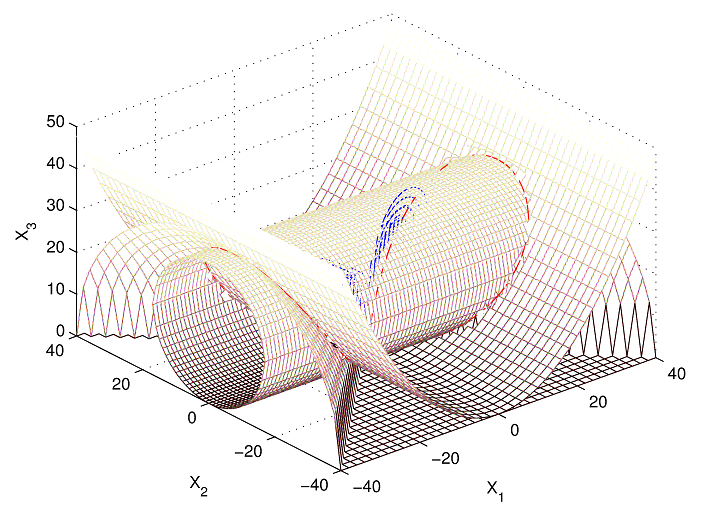}
\end{center}
\caption{For $b \leq 2$ the strange attractor is localized by the corresponding surfaces of the homoclinic orbit. Here for $b=2$ are shown just three of these surfaces. The strange attractor is fairly seen in blue in the interior and the homoclinic orbit is shown in red.}
\label{fig:b2}
\end{figure}

The following surfaces of \eref{local} can be found in Doering and Gibbon \cite{lor:loc2}: 
the repelling parabolic cylinder $-X_1^2+2\rho X_3 = 0$ 
($\lambda \rightarrow -\infty$), the repelling cone $-X_1^2 + X_2^2 + X_3 ^2 =0$ ($\lambda = 0$), the attracting cylinder
$X_2^2 + (X_3-\rho) ^2 =\rho^2$ ($\lambda = 1$) and the attracting sphere $X_1^2 + X_2^2 + (X_3-2\rho) ^2 =(2\rho)^2$ ($\lambda = 2$). The two-sheeted hyperboloids
($\lambda < 1$) (see \Fref{fig:repHypb}) were for the first time found by Giacomini and Neukirch \cite{lor:loc3}.
In \Fref{fig:homA} we see that the repelling cone and the repelling parabolic surface define the homoclinic orbits of the
non-dissipative system. For $b \leq 2$ these homoclinic orbits define also the attracting surfaces (\Fref{fig:b2}).

\section{Correspondence with physical systems}\label{sec:physical} 

	The different Nambu doublets can be interpreted as different physical systems. The one Nambu Hamiltonian is regarded to be the energy of the system and the other has different interpretations depending on the case (a constraint of the motion or an integral of motion). This framework is providing a way to identify completely different physical systems that have the same dynamics. From the mathematical point of view it is just one system in different representations. 
	In our case these are the Heisenberg, $SE(2)$, $SO(3)$ and $SO(2,1)$ formulations that are linked with each other. 

	\subsection{The unforced, undamped Duffing oscillator}\label{sec:duffing}
	Let us work with the first doublet $(H,C)$ of \Tref{tab:doublets}.
We reduce the system on the parabolic cylinder $C = -\frac{1}{2}X_1^2 + \rho X_3$ and the Hamiltonian becomes:
\begin{equation*}
	H = \frac{1}{2}X_2^2 - \frac{1}{2\rho^2}(\rho^2 - C) X_1^2 + \frac{1}{8\rho^2}X_1^4 + const.
\end{equation*}
We see that $\forall \lambda$ the Hamiltonian is the same up to a constant shift. 
\begin{figure}[t!]
\begin{center}
	\includegraphics[width=70mm,height=50mm]{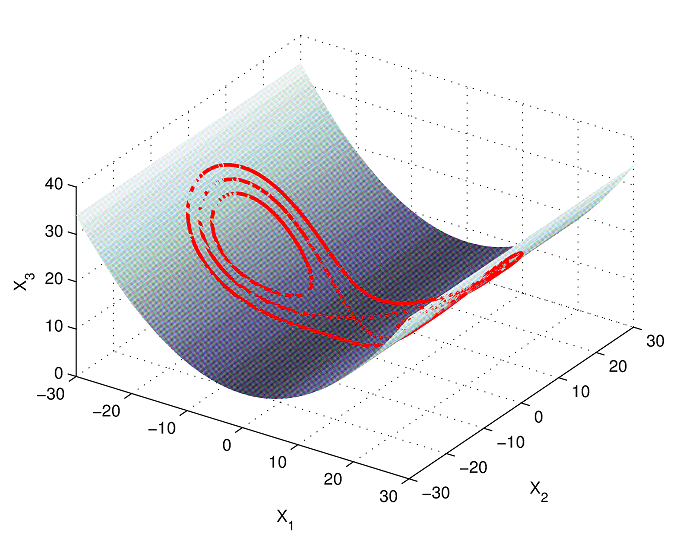}
\end{center}
\caption{The phase space portrait of the system \eref{consLor} reduced on the parabolic cylinder $C$. Choosing initial conditions for the system \eref{consLor}
is equivalent to choosing not only initial conditions but also a potential for the system \eref{duff:ham}. The value of the potential is determined by the level of the surface $C$.}
\label{fig:parPS}
\end{figure}
Let $\rho$ be absorbed in the Hamiltonian, then
\begin{equation}\label{duff:ham}
	H = \frac{1}{2}\rho X_2^2 - \frac{1}{2\rho}(\rho^2 - C) X_1^2 + \frac{1}{8\rho}X_1^4 
\end{equation}
Our dynamical variables are now the conjugate canonical variables $(X_1,X_2)$ and the Nambu bracket reduces to Poisson bracket on a $2D$ phase space (\Fref{fig:parPS}).
\begin{equation*}
	\{X_1,X_2\}_C = 1
\end{equation*}
The system \eref{duff:ham} describes a particle moving along $X_1$ 
with momentum $X_2$ under the influence of the potential:
$
	V(X_1) = -\frac{1}{2\rho}(\rho^2 - C) X_1^2 + \frac{1}{8\rho}X_1^4 
$
where $C=const.$ depends on the initial conditions. The equilibria depend on the value of $C$. They are $\bar{X} = (0,0)$ for  
$C \geq \rho^2$ and 
\begin{equation*}
	\bar{X} = 
\cases{
	(0,0) \\
	(-\sqrt{2(\rho^2 - C}),0) &for $C < \rho^2$ \\
	(\sqrt{2(\rho^2 - C}),0) 
 }
\end{equation*}
\begin{figure}[t!]
\begin{center}
	\subfigure[$C \geq \rho^2$]{\label{fig:2Da}\includegraphics[width=30mm,height=30mm]{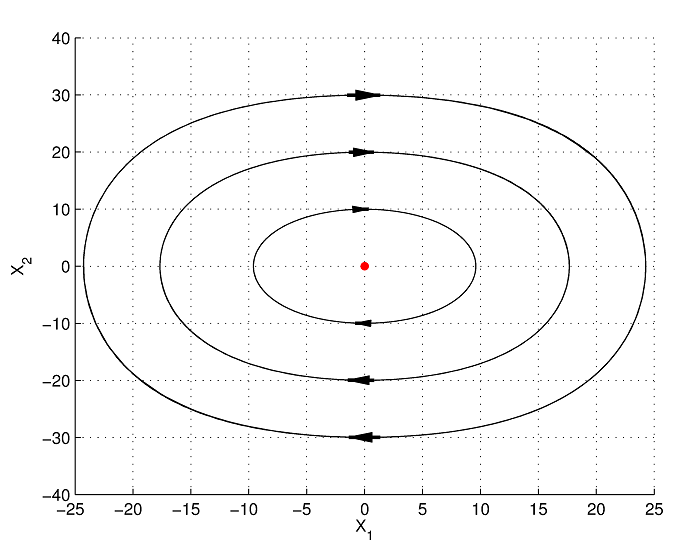}} 
\qquad\qquad
	\subfigure[$C < \rho^2$]{\label{fig:2Db}\includegraphics[width=35mm,height=30mm]{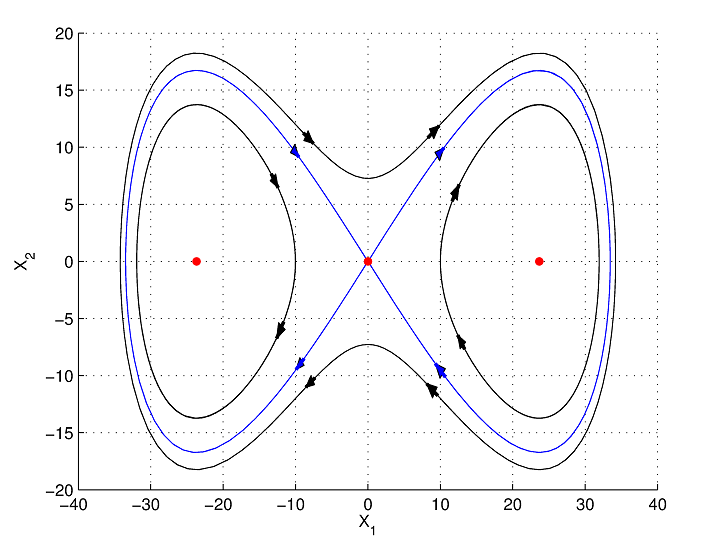}}
\caption{The phase space of system \eref{duff:ham}. It depends on the initial conditions of system \eref{consLor} and specifically on $X_3(t=0)$ through the value of the function $C$.
\label{fig:2D}}
\end{center}
\end{figure} 
This system is the unforced, undamped Duffing oscillator 
$
	\ddot{x} = - \alpha x - \beta x^3
$
for $\alpha = C - \rho^2$ and $\beta = \frac{1}{2}$. 
For $C > \rho^2$ the potential is a single well while for $C < \rho^2$ the potential is a double well and the symmetry of the vacuum state is `spontaneous broken' (\Fref{fig:2D}). 

This simple analogue of a mechanism of `spontaneous symmetry breaking' gives an intuitive picture of the formation of the attractor. 
\begin{figure}[t!]
\begin{center}
	\subfigure[The $3D$ phase space of the non dissipative part \eref{consLor} and the critical parabolic surface.]
		{\label{fig:criC}\includegraphics[width=50mm,height=40mm]{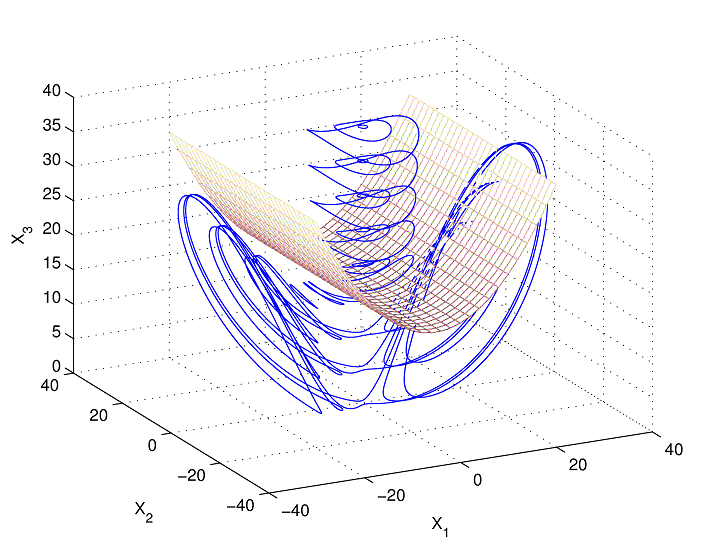}} 
\quad 
	\subfigure[The Lorenz system for initial conditions $\vec{X} = (0,20,40)$ and the critical parabolic surface.]
		{\label{fig:criA}\includegraphics[width=50mm,height=40mm]{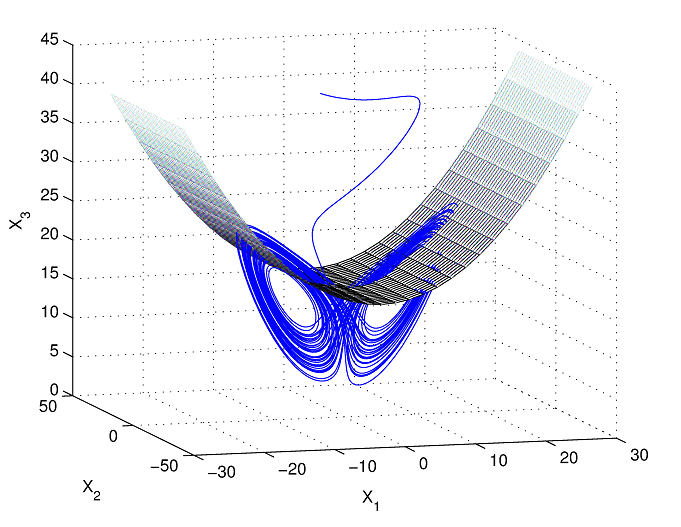}}
\caption{The resemblance with a `spontaneous symmetry breaking mechanism' gives an intuitive explanation for the formation of the attractor.
\label{fig:cri}}
\end{center}
\end{figure} 
Equation $C = \rho^2$ defines a critical parabolic surface that separates phase space in the upper region of one vacuum state and the lower region of two vacuum states (\Fref{fig:criC}). Suppose the initial conditions are above that surface and dissipation is ``turned on". The system will flow towards the origin until it passes the critical surface when it enters the broken symmetry region and has to choose between different vacuum states. It may oscillate around a while but eventually it is forced outside the critical surface again. The procedure is constantly repeated and the attractor is formed (see \Fref{fig:cri}).\\
\indent This is a qualitative picture. However, when the dissipative part is added, the potential becomes another independent variable of the system. With the transformation 
$
	\X_2 = X_2 - \frac{\sigma}{\rho}X_1
$
the full Lorenz system becomes:
\begin{equation*}
\begin{array}{l}                       
	\dot{X}_1 = \rho X_2 \\
	\dot{\X}_2 = -\frac{1}{2\rho}X_1^3 + \frac{1}{\rho} (\rho^2 - \sigma - C) X_1 - (\sigma+1)\X_2 \\
	\dot{C} = (\sigma - \frac{b}{2})X_1 - bC
\end{array}
\end{equation*}
We see that a locally time dependent potential cannot be defined. 
The oscillator develops a friction term and we obtain the Takeyama \cite{takeyama1,takeyama2} memory term in the potential:
$C(t) = C(0) e^{-bt} + (\sigma - \frac{b}{2}) e^{-bt} \int_0^t d\xi e^{b\xi} X_1^2(\xi)$.

	\subsection{Simple pendulum}\label{sec:pendulum}

Now, let work on the Nambu doublet of \Tref{tab:doublets} corresponding to the cylinder 
$
C_{1} = \frac{1}{2}X_2^2 + \frac{1}{2}(X_3 - \rho)^2
$.
Let apply the transformation 
$
	\X_3 = X_3 - \rho
$.
The non dissipative part \eref{consLor} becomes:
\begin{equation}\label{pend:cons}
	\vec{\vv}_{ND} = (\rho X_2 ,
        -X_1 \X_3 ,
		X_1 X_2  )
\end{equation} 
and the full Lorenz system becomes:
\begin{equation}\label{pend:lor}
	\vec{\vv} = (\rho X_2 -\sigma X_1,
        -X_1 \X_3 -X_2,
		X_1 X_2  -b\X_3 - b\rho )
\end{equation} 
The Lorenz system is just translated along $X_3$ axis with the equilibrium of the origin translated at $\X_3 = -\rho$.  Let study independently the volume preserving part \eref{pend:cons}. The cylinder becomes:
$
	C_{1} = \frac{1}{2}X_2^2 +  \frac{1}{2}\X_3^2
$
and we choose the Hamiltonian for $\beta=0$:
\begin{equation}\label{pen:ham}
	H = \frac{1}{2} X_1^2 - \rho \X_3 
\end{equation}
\begin{figure}[t!]
\begin{center}
	\includegraphics[width=70mm,height=50mm]{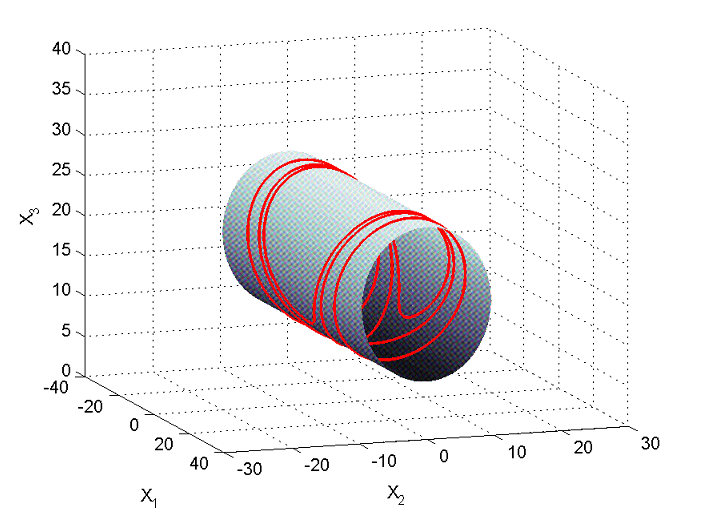}
\end{center}
\caption{The phase space portrait of the system \eref{consLor} reduced on the cylinder $C_{1}$. Choosing initial conditions for the system \eref{consLor}
is equivalent to choosing both initial conditions and the length for a simple pendulum. The length is determined by the radius of the cylinder.}
\label{fig:cylPS}
\end{figure}
Note the analogy of this Hamiltonian to the one of a simple pendulum
$
	H_{pend.} =  \frac{1}{2}p_{\theta}^2/(ml^2) - mglcos\theta 
$.
We perform the following transformations:
\begin{equation*}
	X_1 = -p_{\theta} , \;
	X_2 = R sin\theta ,\;
	\X_3 = R cos\theta 
	\end{equation*}
where our new variables are $(\theta,p_{\theta})$ since:
$
	R = \sqrt{2C_{(1)}} = const.
$
We see that $C_{1}$ is just expressing the constraint that the simple pendulum has constant length. In fact we have parametrized $C_{1}$ with $(\theta,p_{\theta})$: $X_i = X_i(\theta,p_{\theta})$. We have:
$
	\theta = tan^{-1}\frac{X_2}{\X_3}
$
and therefore
\begin{equation*}
\frac{\partial \theta}{\partial X_1} = 0 , \;
\frac{\partial \theta}{\partial X_2} = \frac{\X_3}{2C_1} , \;
\frac{\partial \theta}{\partial \X_3} = - \frac{X_2}{2C_1}
\end{equation*}
Using these derivatives is straightforward to calculate the following Nambu-Poisson brackets:
\begin{equation*}
	\{\theta,p_{\theta}\}_{C_1} = 1 , \;
	\{\theta,H\}_{C_1} =  \frac{\partial H}{\partial p_{\theta}} , \;
	\{p_{\theta},H\}_{C_1} =  -\frac{\partial H}{\partial \theta} 
\end{equation*}
Hence, the canonical equations are retrieved with:
\begin{equation*}
	H = \frac{1}{2}p_{\theta}^2 - \rho R \cos\theta
\end{equation*}
Thus, Lorenz volume preserving motion \eref{pend:cons} reduces to pendulum motion \footnote{
An analogous correspondence between the simple pendulum and the free rigid body can be found in \cite{marsden}
} on level surfaces of $C_{1}$ (\Fref{fig:cylPS}). \\
\indent When the dissipative part is added the equations of motion (full Lorenz system) become:
\begin{equation}\label{pend:cyl}
\begin{array}{l}
        R\dot{\theta} = Rp_{\theta} - (1-b)R \sin\theta \cos\theta  -b\rho \sin\theta \\
        \dot{p_\theta} = -\rho R\sin\theta -\sigma p_\theta \\
		\dot{R} = -b\rho \cos\theta - R(\sin^2\theta + b \cos^2\theta)  
\end{array}
\end{equation}

\subsection{Charged rigid body in a uniform magnetic field}\label{sec:so3}

\begin{figure}[t!]
\begin{center}
	\includegraphics[width=70mm,height=50mm]{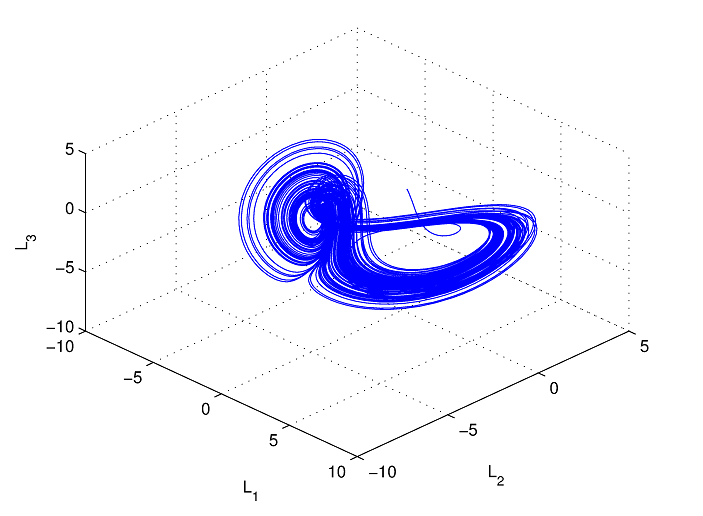}
\end{center}
\caption{The phase-space of system \eref{sph:rsys} for $I_1 = 1/20$, $I_2 = 1/10$, $I_3 = 1/5$, $\rho_1 = 10$, $\rho_2 = \rho_3 = 25$, $\sigma = 10$, $b = 1$ and initial conditions $\vec{L}_o = (-1,1,-1)$.}
\label{fig:rat3}
\end{figure}

Let us work on the Nambu doublet of $C_{2}=  \frac{1}{2} X_1^2 + \frac{1}{2}X_2^2 +  \frac{1}{2}(X_3-2\rho)^2$
with the transformations
\begin{equation}\label{so3tr}
		L_1 = X_1 ,\;
		L_2 = X_2 ,\;
		L_3 = X_3 - 2\rho
\end{equation}
The non dissipative part becomes:
\begin{equation}\label{sph:cons}
	\vec{\vv}_{ND} = (\rho L_2 ,
        -L_1 L_3 - \rho L_1 ,
		L_1 L_2  )
\end{equation} 
and the full Lorenz system becomes:
		\begin{equation}\label{sph:lor}
		\begin{array}{ll}
        \dot{L}_1 = 		\rho L_2 	&-\sigma L_1 \\
        \dot{L}_2 = - L_1L_3 	- \rho L_1	&-L_2 \\
		\dot{L}_3 =  L_1L_2 		&-bL_3 - 2b\rho
		\end{array}	
		\end{equation}		
The Lorenz system is just translated along $X_3$ axis with the equilibrium of the origin translated at $L_3 = -2\rho$. 
The Nambu function $C_{2}$ becomes:
\begin{equation}
	C_{2} = \frac{1}{2} L_1^2 + \frac{1}{2} L_2^2 + \frac{1}{2} L_3^2
\end{equation}
giving the algebra of $SO(3)$. Let work with the corresponding Hamiltonian for $\beta = 1$:
\begin{equation}\label{sph:ham}
	H = L_1^2 + \frac{1}{2}L_2^2 + \frac{1}{2}L_3^2 - \rho L_3 
\end{equation}
We identify $L_i$ as the angular momenta and $C_{2}$ is expressing the conservation of $|\vec{L}|$. This system can be realized as a uniformly charged 
dielectric rigid body with charge $Q$ in a uniform magnetic field $\vec{B} = (0,0,B)$ with moments of inertia along the primary axis $I_1 = 1/2$ and $I_2 = I_3 = 1$ and $\rho = QB$.  
\begin{figure}[t!]
\begin{center}
	\includegraphics[width=70mm,height=50mm]{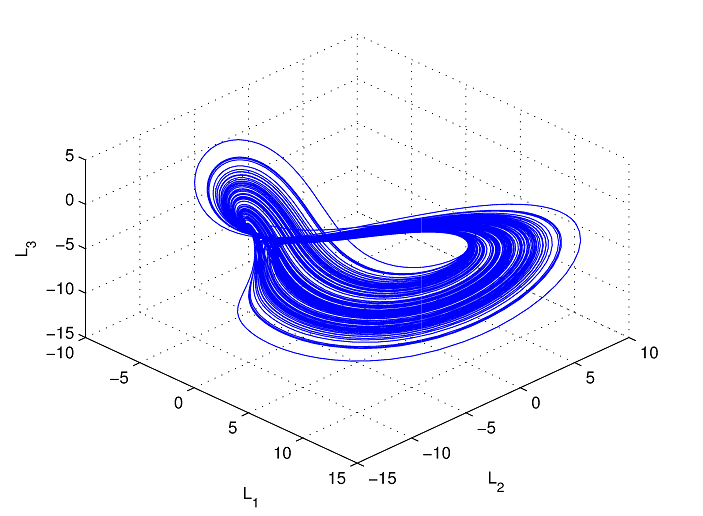}
\end{center}
\caption{The phase-space of system \eref{sph:rsys} for $I_1 = 1/8$, $I_2 = 1/3$, $I_3 = 1/2$, $\rho_1 = 0$, $\rho_2 = 22$, $\rho_3 = 30$, $\sigma = 10$, $b = 1$
and initial conditions $\vec{L}_o = (3,-4,-10)$.}
\label{fig:rat1}
\end{figure}
The dissipative part is just the external torque: a damped term proportional to angular momentum for each component and a constant term $-2b\rho$ in $L_3$ component that acts as a driven torque for $L_3<0$. For an arbitrary rigid body with any moments of inertia and any direction of the magnetic field the Hamiltonian is:
\[
	H = H_{free} + H_{int}
\] 
where
\begin{equation}
\begin{array}{l}
	H_{free} = \frac{1}{2I_1}L_1^2 + \frac{1}{2I_2}L_2^2 + \frac{1}{2I_3}L_3^2 \\[1.2ex]
	H_{int} = - Q\vec{B}\cdot\vec{L}
\end{array}
\end{equation}
Assuming $\vec{\rho} = Q\vec{B}$ and adding the external torque
$
\vec{N}_{ext} = (	-\sigma L_1, -L_2, -bL_3 - 2b\rho)
$
we get the equations of motion for the general system:
\begin{equation}\label{sph:rsys}
\begin{array}{c}
\dot{L}_i = \{L_i , H\}_{C_{2}} + \vec{N}_{ext} \Rightarrow \\
\begin{array}{lll}
        \dot{L}_1 = \left( \frac{1}{I_2} - \frac{1}{I_3}\right) L_2L_3		&+\rho_3 L_2 - \rho_2 L_3	&-\sigma L_1 \\
        \dot{L}_2 = \left( \frac{1}{I_3} - \frac{1}{I_1} \right) L_1L_3 	&+\rho_1 L_3 - \rho_3 L_1	&-L_2 \\
		\dot{L}_3 = \left( \frac{1}{I_1} - \frac{1}{I_2} \right) L_1L_2 	&+\rho_2 L_1 - \rho_1 L_2	&-bL_3 - 2b\rho_3
\end{array}	
\end{array}	
\end{equation}
which can also be written as
$
	\dot{\vec{L}} + \vec{\omega}\times \vec{L} + Q \vec{B}\times \vec{L} = \vec{N}_{ext}
$.
System \eref{sph:rsys} presents strange attractors as can be seen in \Fref{fig:rat1} and \Fref{fig:rat3}.

\subsection{$SO(2,1)$ formulation}\label{sec:so21}

\begin{figure}[t!]
\begin{center}
	\includegraphics[width=70mm,height=48mm]{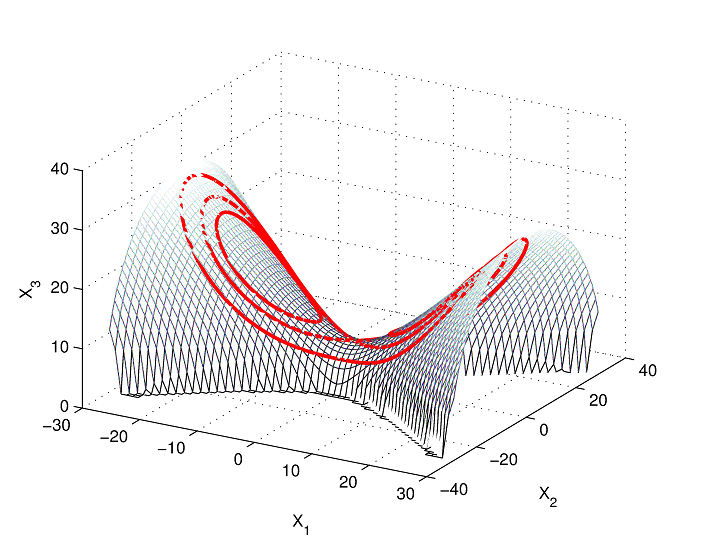}
\end{center}
\caption{The phase-space of system \eref{consLor} reduced on the hyperboloid $C_{0}$ for the case of $AdS_2$.} 
\label{fig:hphsp}
\end{figure}

The $SO(2,1)$ case can be regarded as a `hyperbolic' analogue of the rigid body $SO(3)$ case. This is mainly of mathematical interest. One gets this hyperbolic analogue with the Nambu doublet ($\beta = 2$ for example)
\begin{equation}\label{hyp:sys}
	C_{0}=  -\frac{1}{2} X_1^2 + \frac{1}{2}X_2^2 +  \frac{1}{2}X_3^2 \, , \,
	H=  -\frac{1}{2} X_1^2 + X_2^2 +  X_3^2 - \rho X_3
\end{equation}
Depending on the initial conditions the hyperboloid $C_{0}$ may be two-sheeted which corresponds to the hyperbolic space $\mathbb{H}^2$ or one-sheeted, which corresponds to anti-de Sitter (or de Sitter) space $AdS_2$ (or $dS_2$). The phase space portrait of system \eref{hyp:sys} for the $AdS_2$ case can be seen in \Fref{fig:hphsp}.

\section{Conclusions}

The gauge sector of the phase space geometry of a three dimensional volume preserving dynamical system is investigated. As a toy model we used the non dissipative part of the Lorenz system. Searching for possible valuable informations one may retrieve from this geometric picture, we found that a rich dynamical structure is revealed. 

This rich dynamical structure we are referring to, corresponds to different systems which in our application are the Duffing oscillator (Heisenberg case), the simple pendulum ($SE(2)$ case), a uniformly charged rigid body in a uniform magnetic field ($SO(3)$ case) and a mathematical $SO(2,1)$ formulation. All these turn out to be different representations of the same dynamics! Similar analysis can be performed to any three dimensional volume preserving dynamical system providing insight on a unified description of dynamics for specific systems.

When the full dissipative system is considered the invariant manifolds of the conservative part can be used for the localization of the strange attractor. In addition, in section \ref{sec:duffing}, an elementary analogue of a `spontaneous symmetry breaking mechanism' is proposed as an intuitive explanation for the formation of the attractor. For the $SO(3)$ case, in section \ref{sec:so3}, the Lorenz system is physically identified 
with the rigid body system in a homogeneous magnetic field with external friction in every direction and a constant driven term in $Z$ direction.
A generalization of this rigid body system for arbitrary magnetic field's direction leads to a dynamical system with new strange attractors (figures \ref{fig:rat1} and \ref{fig:rat3}).  

Last, we note that the $SO(3)$ formulation could be useful for an angular momentum or spin-based quantization of the Lorenz attractor in accordance with the matrix formulation of \cite{Axenides:2009rh,Floratos:2011ct}. It may also be interesting to investigate the correspondence with the $SO(2,1)$ picture in this quantum realm.

%%%%%%%%%%%%%%%%%%%%%%%%%%%%%%%%%%%%%%%%%%%%%%%%%%%%%%%%%%%%%

\section*{Acknowledgements}

I am grateful to M. Axenides and E. Floratos for the most useful discussions, suggestions and guidance. This research
has been co-financed by the European Union (European Social Fund - ESF) and Greek national funds through the Operational
Program ``Education and Lifelong Learning" of the National Strategic Reference Framework (NSRF) - Research Funding Program:
THALES. 

\appendix
\section{} \label{app:lag}
Applying the Lagrange's multipliers method, we find the extremum of $S(X_1,X_2,X_3)$ under the constraint $\dot{S}=0$
for: 
\[
	S = (\lambda - 1)X_1^2 + X_2^2 + (X_3-\lambda \rho)^2
\] 
and $(X_1,X_2,X_3)$ subject to the Lorenz flow \eref{lor}. Let call $A \equiv \dot{S}$.

We are looking for a point $(\tilde{X}_1,\tilde{X}_2,\tilde{X}_3)$ and a Lagrange's multiplier $\alpha$ such, that:
$K = S(\tilde{X}_1,\tilde{X}_2,\tilde{X}_3)$ is an extremum of $S$ under the constraint $A=0$.
The constraint gives one equation:
\begin{equation}\label{app:cond1}
	-\sigma (\lambda - 1)\tilde{X}_1^2 - \tilde{X}_2^2 - b \tilde{X}_3^2 + b(\lambda\rho)\tilde{X}_3 = 0
\end{equation}
Let extremize: $f = S + \alpha A$. From: $\nabla f = 0 \Rightarrow \nabla S + \alpha \nabla{A} = 0$ we get three more equations:
\begin{eqnarray}\label{app:cond2}
	\tilde{X}_1 (\lambda - 1)(1-2\alpha \sigma) = 0 \\
\label{app:cond3}	\tilde{X}_2 (1 - 2\lambda) = 0 \\
\label{app:cond4}	\tilde{X}_3 =  \lambda\rho\frac{1-\alpha b}{1-2\alpha b} 
\end{eqnarray}
The system of Equations \eref{app:cond1}-\eref{app:cond4} has three distinct solutions:
\begin{enumerate}
	\item $(\tilde{X}_1,\tilde{X}_2,\tilde{X}_3) = (0,0,0)$ and $\alpha = 1/b$ \\
	which gives $K = S(\tilde{X}_1,\tilde{X}_2,\tilde{X}_3) = (\lambda \rho)^2$ \\
	
	\item $(\tilde{X}_1,\tilde{X}_2,\tilde{X}_3) = (0,\sqrt{(\lambda\rho)^2\frac{b^2(b-2)}{4(b-1)^2}},
			\lambda\rho\frac{2-b}{2(1- b)})$ and $\alpha = 1/2$ \\
	which gives $K = S(\tilde{X}_1,\tilde{X}_2,\tilde{X}_3) = (\lambda \rho)^2 \frac{b^2}{4(b-1)}$. From
	$\tilde{X}_2$ we see the constraint $b>2$. \\
	
	\item $(\tilde{X}_1,\tilde{X}_2,\tilde{X}_3) = (\sqrt{(\lambda\rho)^2\frac{b^2}{\sigma(1-\lambda)}\frac{(2\sigma -b)}{4\sigma(\sigma-b)^2}}
			,0,\lambda\rho\frac{2\sigma-b}{2(\sigma- b)})$ and $\alpha = 1/2\sigma$ \\
	which gives $K = S(\tilde{X}_1,\tilde{X}_2,\tilde{X}_3) = -(\lambda \rho)^2 \frac{b^2}{4\sigma(\sigma - b)}$. From
	$\tilde{X}_1$ we see the constraint $\lambda < 1$. \\
\end{enumerate}

\section{} \label{app:nor}
Let a surface be defined by $S = K$ for some constant $K$ and $S = S(\vec{X})$. If $A = \nabla{S}\cdot\vec{\vv}$ has the same sign at each and every point of $S$ then $S$ is a localization surface. Let calculate $A$ for Lorenz flow $\vec{\vv}$ \eref{lor}
and $S = (\lambda - 1)X_1^2 + X_2^2 + (X_3-\lambda \rho)^2$:
\begin{eqnarray}\label{app:product}
\fl	A \equiv \nabla{S}\cdot\vec{\vv} = (2(\lambda - 1)X_1,2X_2,2(X_3-\lambda\rho)) \nonumber \\
			 \ast (\rho X_2 - \sigma X_1,-X_1X_3 +\rho X_1 - X_2,X_1X_2 - bX_3) \nonumber\\
\lo{=} -2\sigma (\lambda - 1)X_1^2 - 2X_2^2 - 2bX_3^2 + 2b(\lambda\rho)X_3
\end{eqnarray}
We will prove that $A$ has a constant sign for all repelling surfaces of \Eref{local2}.
\begin{itemize}
	\item two sheeted hyperboloids $\lambda < 1$ : $S = -R_r^2 \equiv -(\lambda\rho)^2\frac{b^2}{4\sigma(\sigma - b)}$ \\

	Substituting: $
		(\lambda - 1)X_1^2 = -X_2^2 - X_3^2 + 2(\lambda\rho)X_3 - (\lambda\rho)^2 \frac{(2\sigma - b)^2}{4\sigma(\sigma - b)}
	$ 
	in \eref{app:product} we get:
\begin{eqnarray*}
\fl	A = 2 (\sigma -1)X_2^2 + 2(\sigma -b)X_3^2 - 2(\lambda\rho)X_3(2\sigma - b) 
		  +2\sigma (\lambda\rho)^2\frac{(2\sigma - b)^2}{4\sigma(\sigma - b)} \\
 \lo{=} 2(\sigma - 1)X_2^2 + \frac{1}{2(\sigma - b)} \\
 \times \left\{ 4X_3^2(\sigma - b)^2 - 4(\lambda\rho)(2\sigma - b)X_3(\sigma - b) +(\lambda\rho)^2(2\sigma - b)^2 \right\} \\
\lo{=} 2(\sigma - 1)X_2^2 + \frac{1}{2(\sigma - b)}\left\{ 2X_3(\sigma - b) - (\lambda\rho)(2\sigma - b) \right\}^2 > 0 
\end{eqnarray*}		
$\forall (X_1,X_2,X_3) \in S$ and for $\sigma > 1$, $\sigma > b$. \\

	\item cone $\lambda = 0$ : $S = \tilde{R}_r^2 \equiv (\lambda \rho)^2 = 0$ \\
	
	Substituting: $X_1^2 = X_2^2 + X_3^3$ in \eref{app:product} we get:
	\[
 		A = 2(\sigma - 1) X_2^2 + 2(\sigma - b)X_3^2 > 0 
	\]
	$\forall (X_1,X_2,X_3) \in S$ and for $\sigma > 1$, $\sigma > b$. \\

	\item parabolic surface $\lambda \rightarrow -\infty$ : $S = \tilde{R}_r^2 \equiv (\lambda\rho)^2
				\overset{\lambda \rightarrow -\infty}{\Longrightarrow} -X_1^2 + 2\rho X_3 = 0$ \\
	
	Substituting $\rho X_3$ in \eref{app:product} we get:
\begin{eqnarray*}
\fl	A = (-X_1,0,\rho) 
			 \cdot(\rho X_2 - \sigma X_1,-X_1X_3 +\rho X_1 - X_2, X_1X_2 - bX_3) \\
\lo{=} \sigma X_1^2 - b\rho X_3 = \sigma X_1^2 - \frac{b}{2}X_1^2 = (\sigma - \frac{b}{2}) X_1^2 > 0 
\end{eqnarray*}		
$\forall (X_1,X_2,X_3) \in S$ and for $\sigma > b/2$.	\\

\item one sheeted hyperboloids $\lambda < 0$ : $S = \tilde{R}_r^2 \equiv(\lambda \rho)^2$ \\

	Substituting: $(\lambda - 1)X_1^2 = -X_2^2 - X_3^2 + 2(\lambda\rho)X_3$ in \eref{app:product} we get:
\[
 	2 (\sigma -1)X_2^2 + 2(\sigma -b)X_3^2 - 2(\lambda\rho)X_3(2\sigma - b) > 0
\]
$\forall (X_1,X_2) \in S$ and $X_3>0$ and for $\sigma > 1$, $\sigma > b$. Only the part $X_3>0$ of the surface is 
a repelling surface. From the parabolic surface we know that the attractor lies in $X_3>0$. So these
upper parts of one sheeted hyperboloids provide a lower boundary of the attractor.
\end{itemize}

\section*{References}
% refs.tex                           Symmetries,Lorenz Attractor

\end{document}